\documentclass[preprint]{aastex631}
\newcommand{\nfig}[1]{Figure~\ref{#1}}

\usepackage{amsmath,amssymb}
\usepackage{subfigure}        
\usepackage{txfonts}
\usepackage{sidecap}
\usepackage{multirow}
\usepackage{float}
\usepackage{longtable}
\usepackage{graphicx}
\usepackage{hyperref}
\usepackage{threeparttable}
\usepackage{indentfirst}
\usepackage{natbib}
\usepackage{hyperref}
\usepackage{epstopdf}
\usepackage{footnote}
\usepackage{tabularx}
\usepackage{booktabs}

\usepackage{bm}
\shortauthors{Zhou et al.}
\begin{document}
\title{Successive Coronal Jets as Novel Facilitators for Filament Oscillation and Eruption}

\author{Chengrui Zhou}
\affiliation{School of Physics and Astronomy, Yunnan University, Kunming 650500, People's Republic of China}
\affiliation{Yunnan Key Laboratory of the Solar physics and Space Science, Kunming 650216, People's Republic of China}

\author{Chun Xia}
\affiliation{School of Physics and Astronomy, Yunnan University, Kunming 650500, People's Republic of China}

\author{Wentai Fu}
\affiliation{School of Intelligence Science and Technology, Nanjing University, Suzhou 215163, People's Republic of China}

\author{Hechao Chen}
\affiliation{School of Physics and Astronomy, Yunnan University, Kunming 650500, People's Republic of China}

\author{Qiaoling Li}
\affiliation{School of Physics and Astronomy, Yunnan University, Kunming 650500, People's Republic of China}

\newcommand{\mybold}[1]{\textcolor{red}{{#1}}}
\newcommand{\unit}[1]{\ensuremath{\,\mathrm{#1}}}
\newcommand{\figref}[2]{\nfig{#1}{#2}}
\newcommand{\Alfven}{Alfv\'{e}n\xspace}
\newcommand{\addref}{{\bf REF\xspace}}

\newcommand{\dbr}[1]{\left\llbracket#1\right\rrbracket}
\newcommand{\etheta}{\skew3\hat{\mbox{\boldmath $\theta$}}}

\newcommand{\Cs}{\mathcal{\mathcal C}^{*}}
\newcommand{\Cc}{\mathcal{\mathcal C}}
\newcommand{\Bcom}{{\bm B}_{\mathrm{com}}}
\newcommand{\Bmfr}{{\bm B}_{\mathrm{MFR}}}
\newcommand{\BI}{{\bm B}_{I}}
\newcommand{\BIseg}{{\bm B}_{I_\mathrm{seg}}}
\newcommand{\BIout}{{\bm B}_{I_\mathrm{out}}}
\newcommand{\BF}{{\bm B}_{F}}
\newcommand{\BFseg}{{\bm B}_{F_\mathrm{seg}}}
\newcommand{\BFout}{{\bm B}_{F_\mathrm{out}}}
\newcommand{\Bamb}{{\bm B}_{\mathrm{amb}}}
\newcommand{\Bp}{{\bm B}_{\mathrm{p}}}
\newcommand{\jI}{{\bm j}_{I}}
\newcommand{\jIseg}{{\bm j}_{I_\mathrm{seg}}}
\newcommand{\jIout}{{\bm j}_{I_\mathrm{out}}}
\newcommand{\jF}{{\bm j}_{F}}
\newcommand{\jFseg}{{\bm j}_{F_\mathrm{seg}}}
\newcommand{\jFout}{{\bm j}_{F_\mathrm{out}}}
\newcommand{\Tv}{\hat{\bm T}}
\newcommand{\Nv}{\hat{\bm N}}
\newcommand{\Nvp}{\Nv^{\prime}}
\newcommand{\Mv}{\hat{\bm M}}
\newcommand{\et}{\hat{\bm \theta}}
\newcommand{\ep}{\hat{\bm \phi}}
\newcommand{\ero}{\hat{\bm \rho}}
\newcommand{\eo}{\hat{\bm \omega}}
\newcommand{\kp}{\kappa^{\prime}}
\newcommand{\Rc}{R_{\mathrm{c}}}
\newcommand
{\Ov}
{\bm{\mathcal{O}}}
\newcommand
{\nv}
{\hat{\bm{n}}}

\correspondingauthor{Chun Xia}
\email{chun.xia@ynu.edu.cn}

\begin{abstract}
Solar filament eruptions are central to coronal mass ejections and space weather, yet their triggering mechanisms remain a fundamental open question. In particular, the early-stage that drives a magnetic flux rope toward instability and its observable signatures are poorly understood. Here, combining multi-instrument observations, we report successive coronal jets impacting a filament, causing its gradual rise and oscillations with growing amplitude and period. When the filament reaches the height where the decay index exceeds the torus instability threshold, the rapid filament eruption commences. This filament eruption is reproduced by magnetohydrodynamic simulations, in which successive thermal jets disturb a stable filament in a magnetic flux rope and excite {oscillations} together with the eruption of the filament. As the filament rises to erupt, the restoring forces for the oscillation progressively weaken, which naturally leads to an increase of the oscillation amplitude and period. Our results demonstrate the growing oscillations as one of the observable precursors for filament eruptions, enhancing our ability to predict solar eruptions.
\end{abstract}


\section{Introduction}\label{intro}

Solar filaments are intriguing solar coronal structures, characterized by their low-temperature and high-density plasma suspended within the much hotter solar corona~\citep{1998SoPh..182..107M, 2007ApJ...667L.105J, Mackay2010,2021A&A...647A.112Z}. Their {existence} {mostly} lies along photospheric polarity inversion lines (PILs), indicating highly non-potential magnetic structures such as sheared arcades and flux ropes~\citep{2022ApJ...934L...9L, rust96, zhangj12,2016ApJ...823...22X,2014ApJ...780..130X,2025ApJ...987...38Z}, {which brings} one of the most critical questions for solar physics: what governs the sudden transition from stable suspension to eruption {of solar filament}? This is not merely {an academic question}, filament eruptions are the dominant progenitors of coronal mass ejections (CMEs)~\citep{Munro1979, Webb1987, Gopalswamy2003,2011LRSP....8....1C,2021ApJ...923...45Z,2024ApJ...964..125S,2024ApJ...962...42H,2025ApJ...995..130C}. Therefore, reliable prediction of filament eruptions represents the keystone of space weather forecasting. Achieving this, requires us to identify the precise physical precursor {features to eruption}. 

High-resolution observations now reveal that the dramatic transition from the stable phase to eruption phase of the filament is often preceded by the filament oscillations, including vertical, horizontal transverse oscillations~\citep{2014ApJ...786..151S,Gosain2009,2025ApJ...981..139Y,2008A&A...484..487C} {and longitudinal oscillations~\citep{Bi2014}. Longitudinal filament oscillations occur along the filament magnetic flux rope, whereas transverse oscillations are perpendicular to the spine and are primarily governed by magnetic tension and pressure forces within the filament.} The decayless {longitudinal/transverse} oscillation features associated with filament eruptions have long served as one of the direct {precursors} to the eruption~\citep{Luna2018,2007SoPh..246...89I}. Studying these precursor oscillations offers a unique window into the evolution of magnetic forces of the filament before eruption, which is hard to detected in the corona~\citep{2017ApJ...851...47Z,Tan2023,Mashnich2016,2018ApJ...856..179Z,2012A&A...542A..52Z,2024ApJ...975..280S,2025Univ...11..401G,2025ApJ...995..104D}. However, what causes these pre-eruptive oscillations and how they are physically linked to the eruption remain poorly understood. Bridging this gap is essential for advancing the ability to predict solar eruptions.

 Based on multi-wavelength and multi-angle observations and simulations, we investigate the physical mechanisms driving the growing oscillations and subsequent eruption of an intermediate filament by successive jets. {We demonstrate that successive coronal jets impact the stable filament, exciting growing oscillations and upward motions of the filament. As the filament rise, the decay index of the background field surpasses the torus instability threshold, then triggering rapid eruption of the filament.} Our combined observational analysis and numerical simulations provide the first direct evidence that the growing oscillations of the filament can serve as one of {the} obvious precursors for the filament eruption. This work offers new insight into the pre-eruptive fine structures of solar {filaments} and advances our ability to predict solar eruptions.

\section{Observations and Methods}\label{sec:data}

The filament under study was analyzed using coordinated data from multiple space-based and ground-based instruments, including the full-disk extreme ultraviolet (EUV) images and photospheric magnetograms were provided by the Atmospheric Imaging Assembly~\citep{2012SoPh..275...17L} (AIA; pixel scale 0.6$^{\prime\prime}$, cadence 12\,s) and the Helioseismic and Magnetic Imager~\citep{2012SoPh..275..229S} (HMI; pixel scale 0.5$^{\prime\prime}$, cadence 45\,s for continuum, 12\,min for magnetograms) aboard the Solar Dynamics Observatory (SDO)~\citep{2012SoPh..275....3P}, respectively. Multi-perspective EUV observations from the Extreme Ultra-Violet Imager (EUVI) onboard the Solar Terrestrial Relations Observatory~\citep{2004SPIE.5171..111W} (STEREO) (pixel scale 1.6$^{\prime\prime}$), together with AIA, enabled three-dimensional observations of coronal structures. The associated coronal mass ejection in the outer corona was tracked using the COR2 coronagraph on STEREO/SECCHI~\citep{2008SSRv..136...67H}. {We utilize the full-disk soft X-ray flux measurements} in the 1--8\,\AA\ and 0.5--4\,\AA\ bands from the Geostationary Operational Environmental Satellite~\citep{1994SoPh..154..275G} (GOES), {the high-resolution soft X-ray imaging was provided by the X-Ray Telescope~\citep{2007SoPh..243...63G} (XRT; spatial resolution 3$^{\prime\prime}$) aboard \emph{Hinode}}. Ground-based observations included high-resolution H$\alpha$ observations from the New Vacuum Solar Telescope~\citep{2014RAA....14..705L} (NVST; spatial resolution 0.136$^{\prime\prime}$, cadence 45\,s) and full-disk H$\alpha$ monitoring from the Solar Magnetic Activity Research Telescope~\citep{2004ASPC..325..319U} (SMART; pixel scale 0.6$^{\prime\prime}$, cadence 2\,min). This multi-instrument, multi-perspective dataset enabled us to get a comprehensive analysis of the coronal jets and the kinematic evolution of the filament. 

\section{Observational Results}\label{sec:result}

\begin{figure}
\begin{interactive}{animation}{movie1.mp4}
\centering
\includegraphics[width=0.8\textwidth]{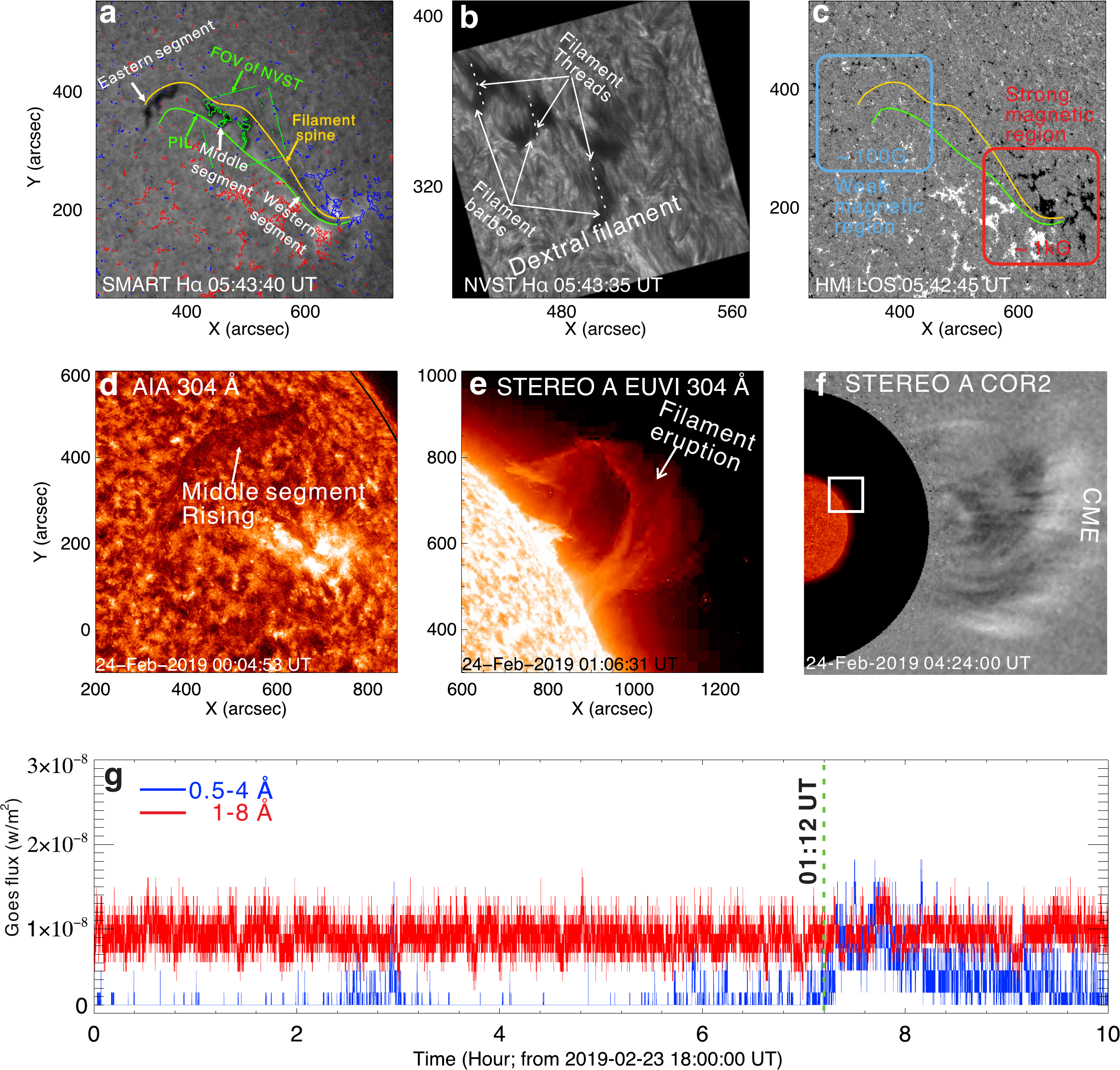}
\end{interactive}
\caption{Overview of the filament and its subsequent eruption. panel (a): SMART H$\alpha$ observation showing the stable filament, the red and blue contours represent positive and negative polarities from the simultaneous line-of-sight magnetogram at levels of $\pm50$~G, respectively. Panel (b): NVST H$\alpha$ observation revealing the filament's fine structure, the white arrows indicate filament threads and barbs, and dashed lines mark the orientation of the barbs. Panel (c): Magnetic environment of the filament. Blue and red boxes delineate regions of relatively weaker ($\sim100$~G) and stronger ($\sim1000$~G) magnetic fields, respectively. Panel (d): The erupting middle filament segment captured in AIA 304~\AA{} during the early phase of eruption. panel (e): Erupting filament observed from the STEREO perspective. panel (f): Running difference images of a CME observed by STEREO/COR2 white-light coronagraph, with STEREO EUVI 304~\AA~image. panel (g): GOES soft X-ray flux profiles: red and blue curves show 1--8~\AA{} and 0.5--4~\AA{} fluxes, respectively.
\\ (An animated version of this figure is available, the animation lasts 6~s and presents the oscillatory filament eruption in the FOV of panel~(d) in AIA 171~\AA{}, 304~\AA{}, and 193~\AA{}, together with XRT soft X-ray images, and also shows the subsequent eruption and CME from the STEREO limb perspective.)}
\label{fig1}
\end{figure}

On 23 February 2019 during solar minimum, a C-shape filament was observed in the Northern Hemisphere (green and yellow curve in \nfig{fig1} a) by the SMART H$\alpha$ observatory. The  photospheric magnetic field underneath the filament (\nfig{fig1} (c)) was highly asymmetric, with diffuse, weak fields ($\sim$100~G) {in} northeast solar quiet region, and concentrated, strong fields ($\sim$1000~G) in a southwest decaying active region. In H$\alpha$ observation (\nfig{fig1} (a)), the filament consists of three distinct segments (white arrows in \nfig{fig1} (a)). The eastern and middle segments of the filament were located in the relatively quiescent region, whereas the western filament segment was extended into the active region. High-resolution NVST H$\alpha$ observation (\nfig{fig1} (b)) shows several threads and barbs in the filament's middle segment, with barbs and threads oriented rightward from the filament spine, indicating a dextral filament chirality with negative magnetic helicity~\citep{2020RAA....20..166C,1998ASPC..150..419M}, consistent with the hemispheric rule. 

Approximately 16~hours later, the middle {filament} segment began to rise at 00:04 UT on 24 February (see \nfig{fig1} (d), {and see movie1}), and evolved into a full eruption observed by STEREO A/EUVI 304~\AA~at 01:36 UT (\nfig{fig1} (e)). Then a clear coronal mass ejection (CME) was subsequently detected by STEREO A/COR2 at 04:24 UT (\nfig{fig1} (f), 
{and see movie1}). Notably, the eruption produced very weak soft X-ray emission: GOES fluxes (\nfig{fig1} (g)) in the 0.5--4~\AA{} channel remained well below the A-class flare, with no discernible enhancement in the 1--8~\AA{} band. Below, we analyze the origin of successive jets, their interaction with the filament, the induced transverse filament oscillations, and the subsequent filament eruption.

\begin{figure*} 
\begin{interactive}{animation}{movie2.mp4}
\centering
\includegraphics[width=0.85\textwidth]{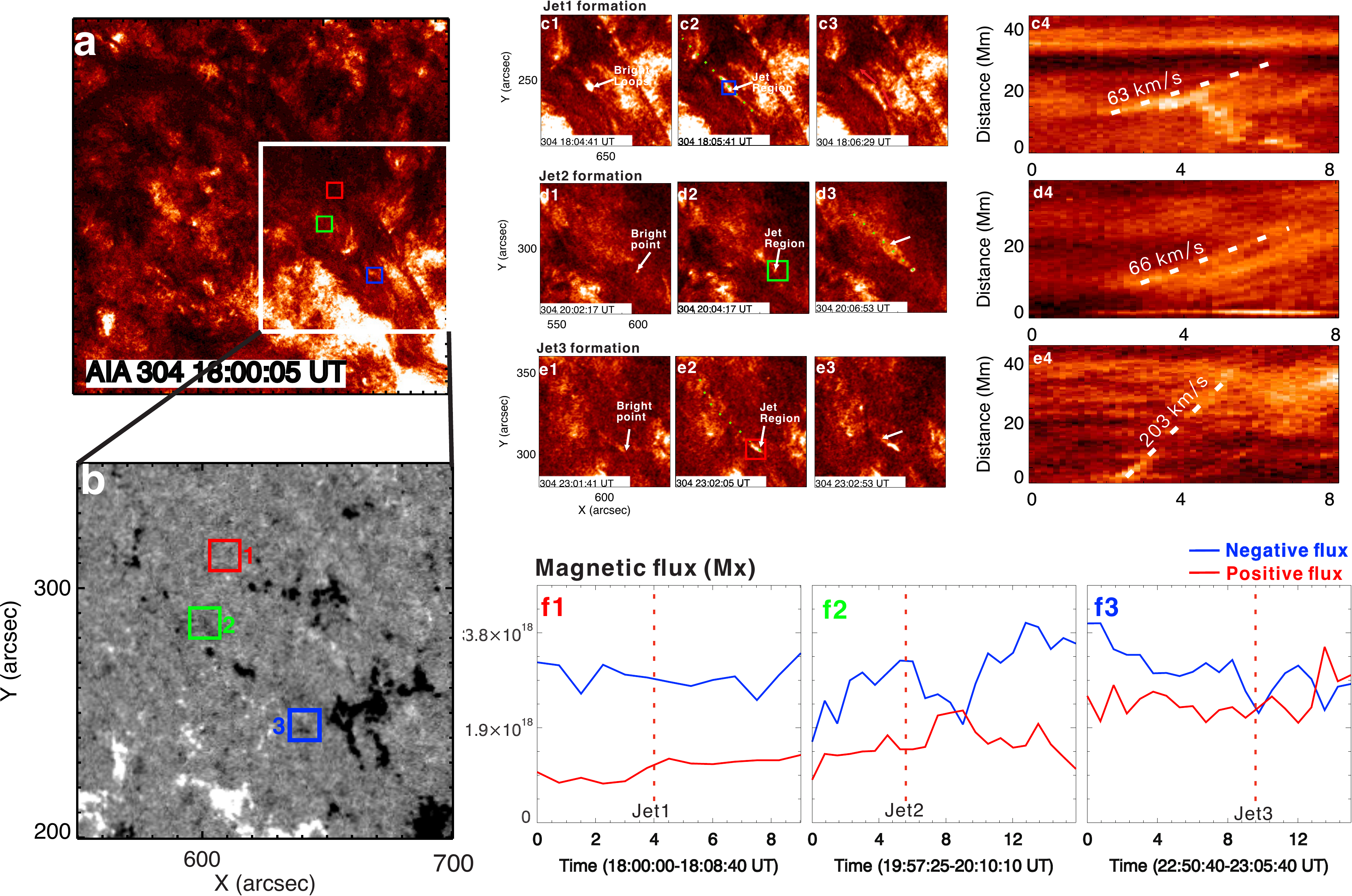}
\end{interactive}
\caption{Triggering process and magnetic evolution for the jet events. Panel (a): Filament observed in the 304~\AA{} channel. Panel (b): Simultaneous magnetogram of the same region from HMI. Panels (c)--(e) show time sequences of AIA 304~\AA~images of the jet triggering process, with panels (c4), (d4), and (e4) displaying the respective time--distance diagrams for each jet. Panels (f1)–(f3) present the time evolution of magnetic flux within the colored boxes in panel (b) (same color label indicates corresponding boxes), as derived from HMI LOS magnetograms, the red and blue curves represent the positive and negative magnetic fluxes, respectively.
\\ (An animated version of this figure is available, the animation lasts 5~s and presents the triggering process of successive jets in the AIA 171~\AA{} and 304~\AA{} observations, with the FOV consistent with panels (c1), (d1), and (e1). It provides a dynamic view of the jet evolution and the magnetic reconnection.)}
\label{fig2}
\end{figure*}

AIA observations reveal that the filament was activated by three successive coronal jets originating beneath {the filament}{~(detailed evolution is shown in movie2)}. As shown in \nfig{fig2}, all three jets originated in the weak negative-polarity regions near the filament's footpoints (\nfig{fig2} (b)). Their evolution followed a similar pattern: each initially appeared as a compact brightening in AIA 304~\AA{} images before expanding into a collimated outflow. Jet~1 began at 18:04 UT, it first appeared as a bright loop, which later developed into a bidirectional outflow; Jets~2 (began at 20:32 UT) and~3 (began at 23:01 UT)  both showed a brightening point and {unidirectional bright flows toward northeast}. {From time–distance diagrams along the jet propagation paths (see the green dotted lines in panels (c2), (d3) and (e3)), we measured the speeds of the three successive jets: $\sim 63$~km~s$^{-1}$ for Jet~1, $\sim 66$~km~s$^{-1}$ for Jet~2, and $\sim 203$~km~s$^{-1}$ for Jet~3.} Classic coronal jets are typically described as transient, collimated ejections with a lasting bright base~\citep{1992PASJ...44L.173S,1995Natur.375...42Y,2025ApJ...985L..11T,2021ApJ...912L..15T}, and often associated with magnetic reconnection between {an} emerging bipole or eruptive mini-filament with the ambient open fields~\citep{2010ApJ...720..757M,2015Natur.523..437S,2019ApJ...883..104S,2025ApJ...987L..21Z,2024ApJ...962L..38D}. However, the lasting bright jet base was not observed in our event. Magnetic flux analysis of the region encompassing the jets (colored boxes in \nfig{fig2} (b)) also shows no significant {cancellation, emergence, or movement of magnetic flux on photosphere} during this period (see \nfig{fig2} (f1)-(f3)). Thus, the jets were possibly triggered by coronal reconnection between the filament's magnetic structure with adjacent coronal loops~\citep{2021NatAs...5...54A,2025ApJ...995...94C} or {the large‑angle reconnection in the lower solar atmosphere that AIA can not detect~\citep{2024A&A...686A.218N,2026A&A...706A...1D}.}

\begin{figure*} 
\centering
\includegraphics[width=0.85\textwidth]{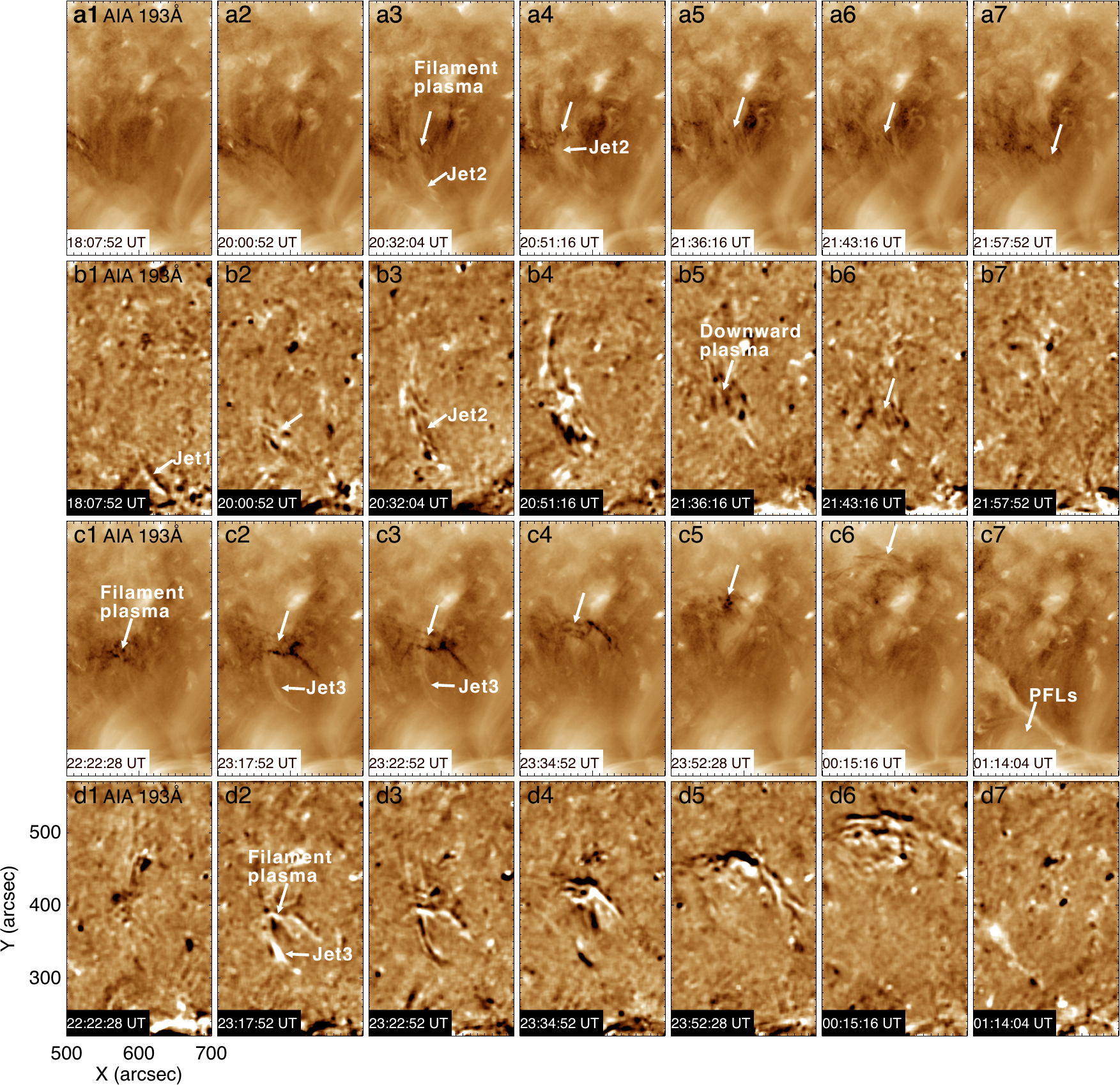}
\caption{Temporal evolution of the filament eruption driven by jet activity. Panels (a1)--(a6) and (c1)--(c6) show AIA 193~\AA{} base-ratio images, capturing the jet activity, filament oscillations, and the formation of post-flare loops during the sequence. Panels (b1)--(b6) and (d1)--(d6) present the corresponding running-difference images in the AIA 193~\AA{} passband, highlighting filament changes and propagating disturbances.}
\label{fig3}
\end{figure*}

\begin{figure}
\centering
\includegraphics[width=0.8\textwidth]{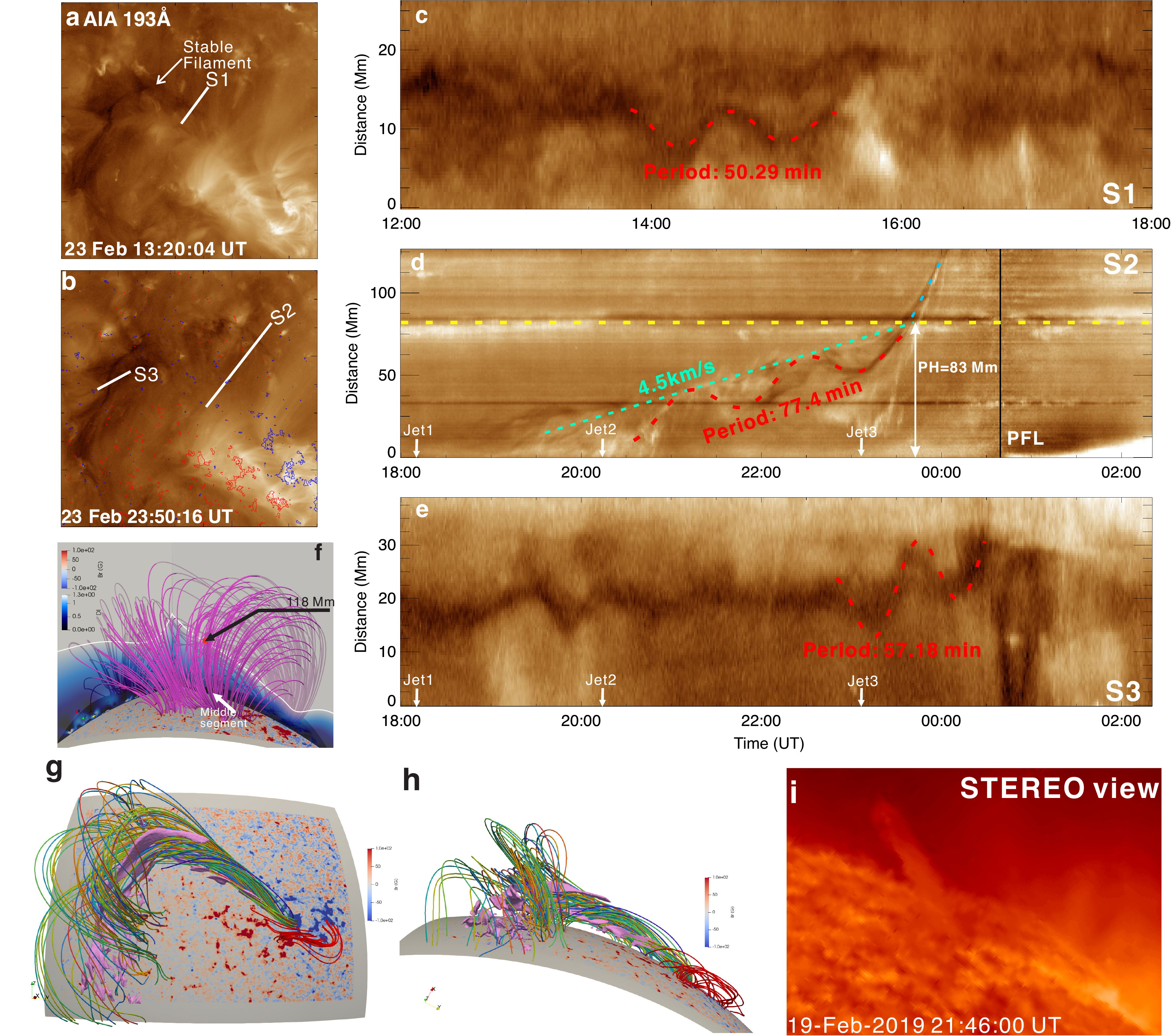}
\caption{Filament kinematics from pre-eruptive oscillations to eruption, and coronal magnetic field from extrapolation and simulation. Panel (a): AIA 193~\AA~images showing the stable filament before Jet~1 impacted. panel (b): AIA 193~\AA~images capturing the eruptive phases of the filament. Panel (c): Time–distance diagram along slice S1 (marked in panel a), revealing the transverse oscillations of the filament before Jet~1 impacts. Panel (d, e): Time–distance diagrams along slits S2 and S3 (positions marked in panel b), showing the growing oscillations of the middle filament segment and the eruption kinematics of the filament. Panel (f): Potential field extrapolation of the coronal magnetic field based on the photospheric magnetogram, purple flux tubes delineate the background field lines above the filament. Panel (g, h): Reconstructed flux rope from the MF simulation in two perspectives. Colored field lines represent the filament flux rope; pink contours outline magnetic dips where filament plasma is trapped. Panel (i): STEREO EUV observation of the filament.}
\label{fig4}
\end{figure}

{The favorable viewing angle and orientation of the filament allow us to observe the filament and its underlying activity simultaneously, while also providing a clear perspective for detecting vertical transverse oscillations with reduced projection ambiguity.} We analyzed the filament dynamics in response to successive jets using AIA 193~\AA{} images and running-difference maps (shown in \nfig{fig3}). Jet~1 interacted with the filament at 18:07 UT, inducing a slight upward motion and {contributing to filament oscillations.} Jet~2 subsequently drove a gradual ascent, enhancing the oscillatory motions by 21:36 UT. Finally, Jet~3 terminated the oscillatory phase by impacting the filament at 23:17 UT, which facilitated the filament's rise. Subsequently, post-flare loops (PFL) appeared, and the filament began its rapid eruption at 01:14 UT. Below, we present observational results of the filament's pre-eruption phase in \nfig{fig4}.

{We note that some oscillations in the middle filament segment existed prior to Jet~1, these pre‑existing oscillations are attributed to thermal pressure disturbance of the decay active region and gentle magnetic reconnection occurring in the eastern filament segment. To capture how filament oscillations evolved toward eruption, we constructed time–slice diagrams comparing the oscillations in three regions of the filament. For the middle segment of the filament oscillations before Jet~1 (prior to 18:00 UT), slice S1 (\nfig{fig4} (a)) was placed perpendicular to the filament's PIL. The transverse oscillations of the stable filament are shown in the time–distance diagram (\nfig{fig4} (c)) obtained from slice S1. After successive jet impacts (after 18:00 UT), the filament began to rise and experience the oscillatory eruption. Slice S2 (\nfig{fig4} (b)) was aligned with the solar radial direction and perpendicular to the filament PIL, the slice S2 also captured the most violent part of the filament eruption. Its time–distance diagram shows the rising filament exhibited transverse oscillations that align closely with jet timings (\nfig{fig4} (d)). In addition, for the stable eastern quiescent filament segment, we constructed a time–slice diagram along slice S3 (\nfig{fig4} (b)), oriented along the filament threads (the direction of the magnetic dips), to capture longitudinal oscillations.} To characterize the large-scale oscillatory behavior of the filament, we fitted the plasma motions to a damped harmonic oscillator model. The fitting equation is $x(t) = A_0 e^{-b t} \cos(p t + \phi_1) + c t + d$, where $A_0$ is the initial amplitude, $b$ the damping rate, $p$ the oscillation frequency, $\phi_1$ the phase, and the terms $c t + d$ account for a linear background drift during the filament's ascent. The model well fits the observed amplitude and period of the transverse oscillations. After Jet~1 and Jet~2 impact, the oscillation exhibited a period of 77.4~min and an amplitude of 12.56~Mm, exceeding those of oscillations of the middle segment before Jet~1 (period 50.29~min, amplitude 4.11~Mm, see \nfig{fig4} (c)). {Notably, longitudinal oscillations measured from the eastern filament threads also show a pronounced increase in amplitude before the filament eruption (period 57.18~min, amplitude 7.32~Mm) (see \nfig{fig4} (e)), consistent with the statistical trend reported by \citet{Luna2018}.} The observed growth in amplitude and period of the transverse and longitudinal oscillation may indicate a progressive reduction in the restoring force during filament ascent. Subsequently, Jet~3 impacted the filament and facilitated its eruption.

From AIA 193~\AA~time-slice diagram of the radial slice S2 (\nfig{fig4} (b)), we derive the height evolution of the filament material. The filament first rose slowly at a speed of 4.5~km~s$^{-1}$ by jet impacts, then underwent a sharp transition to rapid eruption (the blue dashed line in \nfig{fig4} (d)). {Because the filament was located behind the solar disk in the field-of-view (FOV) of STEREO‑A spacecraft during its slow‑rise phase, we could not get its height through the 3D reconstruction by SDO and STEREO-A.} {the projected height is corrected by assuming that the eruption is along the radial direction, and the foot point of the radial slice S2 on the solar sphere, determined by the centroid of the post-flare loops (longitude $\theta=41^\circ$, latitude $\phi=20^\circ$), yields the corrected height via $h = \frac{L_{\text{s2}}}{\sqrt{\sin^2\phi + \cos^2\phi \sin^2\theta}}$. This gives a corrected height of 118~Mm from the projection height of 83~Mm, with an uncertainty of $\pm 3$~Mm. The decay index is defined as  $n = -d\ln|B|/d\ln|h|$~\citep{2006PhRvL..96y5002K}}, the critical decay index for triggering torus instability in filament eruptions is typically in the range 1.1–1.3~\citep{2010ApJ...718.1388D}, with torus instability dominating the rapid eruption phase~\citep{2024ApJ...966...70X}. For reference, we adopt the maximum value of 1.3. According to the potential field extrapolation result (see \nfig{fig4} (f)), the decay index $n$ of the background field above the middle segment reaches 1.3 at 118~Mm. The time–distance diagram shows that the slow rise of the filament material terminates at the true height of 118~Mm, which coincides with the height where the decay index reaches the reference value. Since the flux rope center lies above this altitude, its decay index must surpass the torus instability threshold of 1.3, thereby triggering the rapid eruption. Therefore, the torus instability could be responsible for the rapid eruption phase of the filament.

While torus instability provides a compelling explanation for the observed eruption dynamics, we also considered the potential role of kink instability. The kink instability typically occurs in highly twisted flux ropes (with twist exceeding $\sim$2.0 turns), converting internal magnetic twist into a writhed helical structure, which stores and releases eruptive energy \citep{2005ApJ...630L..97T}. Using the MPI-AMRVAC code \citep{2018ApJS..234...30X, 2023A&A...673A..66K}, we performed a data-driven, three-dimensional magnetofrictional (MF) simulation to reconstruct the filament flux rope via helicity condensation \citep{2013ApJ...772...72A,2022ApJ...934L...9L,2024ApJ...965..160C,2026SCPMA..6949611C} from 17:00 UT on 20 February to 17:00 UT on 23 February 2019. {The horizontal electric field $\mathbf{E}_h$ was inverted from HMI vector magnetograms using the PDFI\_SS code \citep{2020ApJS..248....2F} and was applied as the driving boundary condition.} Following \citet{MCheng2012}, we add a non-inductive electric field $\mathbf{E}_h^{\,n}$ satisfying $\nabla_h \cdot \mathbf{E}_h^{\,n} = \nabla_h^2 \phi = \Omega \, B_r(\theta, \varphi)$, where $\Omega = -1/3$ turn per day to inject negative magnetic helicity from photospheric magnetic elements with strengths exceeding 30~G. The simulation (\nfig{fig4} (g, h)) produces a magnetic flux rope (colored lines), with magnetic dips (pink contours) that are able to contain the filament plasma. These features closely match observations from both SDO and STEREO perspectives (see \nfig{fig4} (a) and (i)). In the flux rope, magnetic twist remains below the kink threshold ($<2$ turns), and no clear helical structures were identified during the eruptive phase in our observations (see \nfig{fig1} (e,f)); thus, we conclude that torus instability rather than kink instability was the dominant driver of the eruption. {It is worth noting that a full MHD simulation with this magnetic field configuration fails, because residual Lorentz forces in quiet-Sun regions near the photosphere are too large and cannot be properly removed with the current MF model.}


\section{numerical model result}\label{sec:discussion}

\begin{figure}
\begin{interactive}{animation}{movie5.mp4}
\centering
\includegraphics[width=0.95\textwidth]{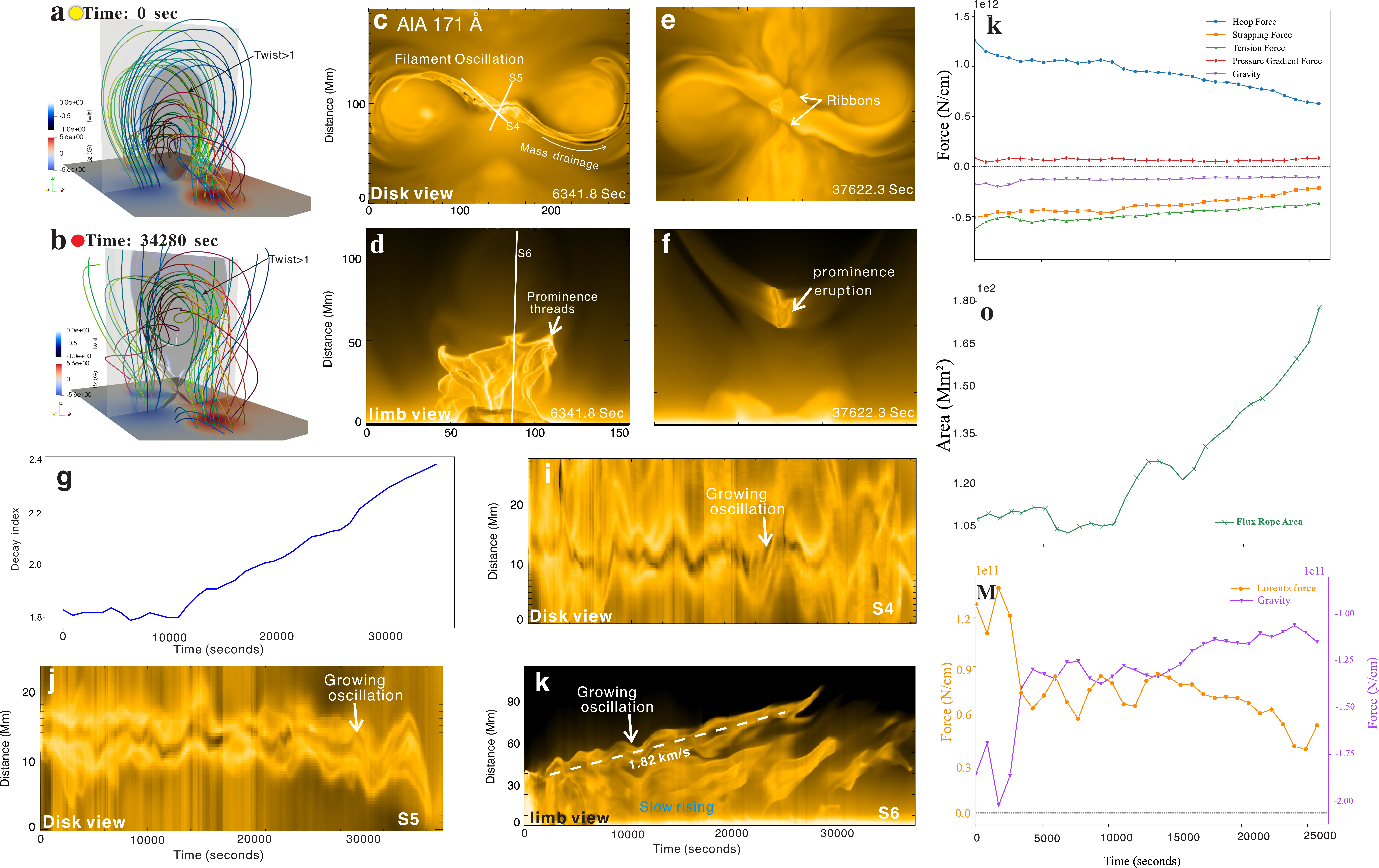}
\end{interactive}
\caption{MHD simulation results for the erupting oscillatory filament, including the height–time profile, growing transverse oscillations, decay index evolution, and kinematic/force analysis of the flux rope. Panels (a) and (b) show the magnetic flux rope in its stable and eruptive configurations, respectively. Panels (c)–(f) present synthetic AIA~171~\AA~images of the filament/prominence from two viewing angles, illustrating both the stable and eruptive phases. Panel (g) displays the evolution of the decay index of the background field as the filament ascends. Panels (i) and (j) show time–slice diagrams along the white lines S4 and S5 in panel (c), respectively. Panel (k) presents the time–slice diagram along the white line S5 in panel (d). Panel (l) shows the forces derived from the MHD simulation, including the Lorentz force components (hoop, strapping, and tension forces), gravity, and the plasma pressure gradient force. Panel (o) shows the temporal evolution of the flux rope cross‑sectional area during the eruption. Panel (m) provides the time profiles of the Lorentz force and gravity acting on the flux rope.
\\(A 12 seconds animated version of this figure is available. The animation presents the full time evolution of the simulated oscillatory filament eruption, including synthetic 171~\AA~images from both disk and limb views, and the magnetic field lines of the flux rope in 2D and 3D perspectives, as well as the transverse and longitudinal filament oscillations, mass drainage, intermittent reconnection, and the eventual eruption.)}
\label{fig5}
\end{figure}

To validate {the} oscillatory eruption of the filament, we performed three-dimensional magnetohydrodynamic (MHD) simulations using the {MPI-}AMRVAC 3.0 code \citep{2023A&A...673A..66K,2018ApJS..234...30X}. Our model incorporates gravity, radiative loss, thermal conduction, and empirical coronal heating. {The synthesis of the EUV 171~\AA~channel follows the method of \citet{2026ApJ..1004..166F}, which includes the absorption lines for optically thick radiation}. The modeled atmosphere comprises a 2,000~km thick chromosphere and a corona extending to 200~Mm. \nfig{fig5} (a) presents the {stable} 3D magnetic flux-rope, which is constructed using Regularized Biot-Savart Laws (RBSL) law~\citep{2021ApJS..255....9T}, the average twist of the flux rope is about 1.7 turns. The flux rope is embedded in an idealized bipolar background field \(B_q\), generated by two magnetic charges of strength \(q = 250\,\mathrm{G}\) located \(80\,\mathrm{Mm}\) below the photosphere and separated by \(30\,\mathrm{Mm}\). Since prominence plasma predominantly accumulates in concave‑up magnetic dips of the prominence magnetic flux rope, we insert the prominence plasma density $\rho_{prom}=2.81\times10^{-14}\, \mathrm{g/cm^{3}}$ to the magnetic dips. After a 3.6-hours relaxation, the simulated filament reaches a stable state. To study the excitation and growth of filament oscillations, we add a localized periodic heating source at the flux rope footpoint~\citep{1999ApJ...512..985A,2022A&A...663A..31N}. The heating rate is $H_l(s,t)=M(t)\cdot10^{-3}e^{-(s-s_{\text{peak}})^2}|\sin(\pi t/600)|$ (cgs), with $M(t)=H(600-\bmod(t,2400))$, $s_{\text{peak}}=2.1$~Mm. It should be noted that the $H(x)$ is the Heaviside step function ($H(x)=1$ for $x>0$, else $0$), which produces successive jets, each lasting 600~s with a 1800~s quiescent interval. The simulation domain spans $[-100,100]\times[-160,160]\times[0,200]~\mathrm{Mm}$ with $390\times416\times390~\mathrm{km}$ resolution. At the bottom boundary, all variables are fixed, and the other boundaries, the no‑inflow open conditions are used.

To analyze the kinematic evolution, we simulated the coupled system for 9.5~hours, capturing its gradual upward motions, {horizontal and vertical} transverse oscillations, and eventual eruption. The eruptive magnetic configurations of the simulated filament are shown in \nfig{fig5} (b). Because the filament looks different when viewed along versus perpendicular to its axis, we adopt two viewing angles to produce synthetic 171~\AA{} images in \nfig{fig5} (c,d) to match the observational perspective. Specifically, \nfig{fig5} (c) presents a view perpendicular to the filament axis, while \nfig{fig5} (d) shows a view along it. {We used three slices to characterize oscillations in different directions to make time-slice diagrams (\nfig{fig5} (i)-(k)): the white line labeled S4 in \nfig{fig5} (c), which is parallel to the magnetic field and captures filament longitudinal oscillations; another white line S5 in \nfig{fig5} (c) that is perpendicular to the magnetic field, captures horizontal transverse oscillations; and the white vertical line S6 in \nfig{fig5} (d), captures vertical transverse oscillation. Our model successfully reproduces key observational features of the filament's pre-eruptive phase: (i) The increasing filament longitudinal oscillations prior to the eruption (see \nfig{fig5} (i)), similar to the on‑disk filament pre‑eruption oscillations of our slice S3 (see \nfig{fig4} (e)); (ii) The growth of the horizontal transverse oscillations in both amplitude and period, which is displayed by the time–distance diagram (\nfig{fig5} (j)) from slice S5 (\nfig{fig5} (c)), it should be noted that the horizontal transverse oscillations were not identified in our observations, which is attributed to the near‑limb viewing perspective of SDO that makes such oscillations difficult to detect;  (iii) The growth of vertical transverse oscillations with the rising motion of the prominence (\nfig{fig5} (k)), from the rising trajectory of filament plasma (white dashed line in \nfig{fig5} (k)), we derive the ascent speed of \(v_{\mathrm{fil}} \approx 1.82~\mathrm{km~s^{-1}}\), these phenomena are consistent with those observed in the pre‑eruption slice S2; } As the filament rises progressively, the decay index of the background field increases accordingly (\nfig{fig5} (g)). Once it surpasses the torus instability threshold, the rapid eruption ensues.

The resulting time–distance diagram reveals a clear pattern: the oscillation amplitude progressively increases as the filament approaches eruption. This amplitude growth is consistent with previous statistical studies identifying filament oscillations as reliable precursors to filament eruptions \citep{2008A&A...484..487C,Luna2018}. {We quantified the forces acting on the filament by integrating the force over the cross-sectional area of the flux rope where the absolute magnetic twist of the magnetic line was greater than unity (see the blue-white contour in \nfig{fig5} (a) and (b)), including tension force, strapping force, hoop force \citep{2015Natur.528..526M, 2021NatCo..12.2734Z}, gravity, and the plasma pressure gradient force.} As the filament rises, the forces (the colored lines in \nfig{fig5} (k)), particularly the Lorentz force within the flux rope, decrease significantly (see the orange line in \nfig{fig5} (m)), while the flux rope gradually expands (see the {green} line in \nfig{fig5} (o)). Simultaneously, as jets drive the flux rope to rise and expand, the gravitational potential wells formed by the filament's magnetic dips become progressively shallower, {causing the trapped plasma to oscillate and drain away from the filament footpoints, which can be seen by comparing the filament material in panels (d) and (f) (a more detailed evolution is available in movie3).} The mass drainage leads to a steady decrease in the gravitational force acting on the filament. Therefore, the Lorentz force and gravity force that govern transverse oscillations are both weakened (see the {purple} line in \nfig{fig5} (m)). {Notably, gravity in the flux rope is about twice as strong as the Lorentz force of the flux rope in \nfig{fig5} (m), and the two forces decrease by comparable magnitudes, indicating that gravity and the Lorentz force both plays a non‑negligible role in the growth of prominence vertical oscillation}. Considering that the oscillatory energy remaining approximately constant, the reduced restoring forces naturally lead to larger period and amplitude, like a spring oscillator whose spring constant is decreasing. The phenomena reported here are not isolated; similar pre‑eruptive oscillatory signatures have been documented in some previous studies \citep{2025ApJ...983..143C,Teng2024,Bi2014,2024A&A...691A.354L}. Thus, the oscillation grows progressively as the filament flux rope approaches MHD instability. Such growing oscillations serve as a precursor for filament eruption driven by torus instability. 

\begin{figure}
\centering
\includegraphics[width=1\textwidth]{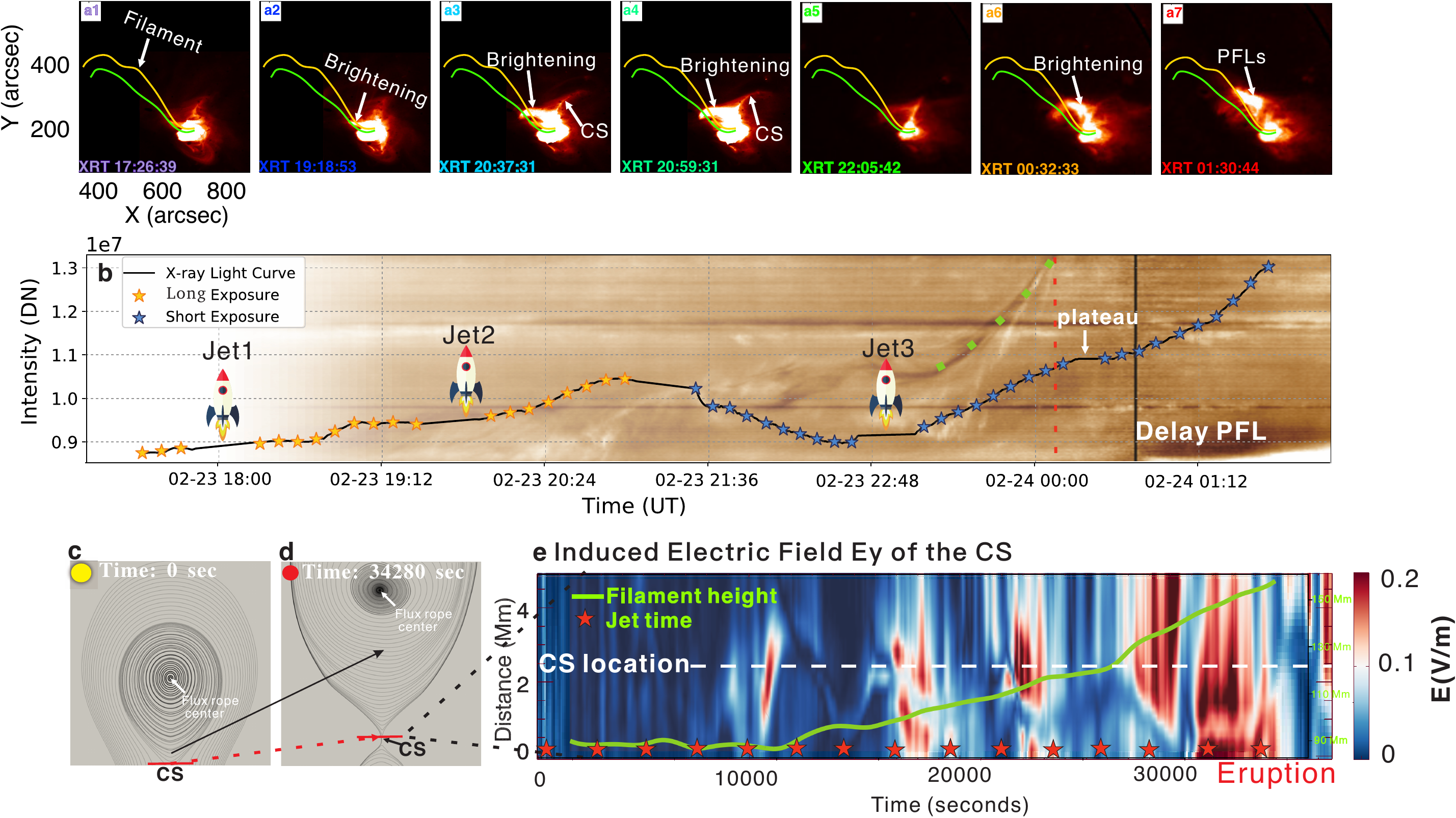}
\caption{Panels (a1)–(a7) present Hinode/XRT soft X‑ray images (Al‑poly filter). Panel (b) shows the soft X‑ray light curve integrated over the FOV of (a1)–(a7); the yellow and blue stars indicate the two exposure modes of XRT (long and short, respectively). Panels (c) and (d) display the two‑dimensional magnetic field lines in the central x–z cross‑section from the simulation. Panel (e) shows the electric field $E_y$ along the current sheet during the filament eruption; the red stars mark the timing of the jets in the simulation, and the filament height is derived from the center of the simulated flux rope.}
\label{fig6}
\end{figure}

{To investigate the role of magnetic reconnection in the rapid-rise phase of the eruption, we combined Hinode/XRT soft X‑ray observations (see \nfig{fig6} (a)) with AIA 193~\AA~time--distance image (see \nfig{fig6} (b)). \nfig{fig6} (a1)-(a7) show the XRT soft X‑ray images in Al‑poly filter. \nfig{fig6} (a2) shows small soft X‑ray brightenings appearing after Jet~1 impacted the filament. As the filament rose, these brightenings subsequently extended and developed into post-flare loops (shown in \nfig{fig6} (a3)--(a7)). We calculated the soft X‑ray flux from the FOV of panels (a1)–(a7), with the time–distance image of slice S2 as the background in panel (b). As the filament rises under jet impacts, the reconnection-associated soft X‑ray emission (blue star curve in \nfig{fig6} (b)) does not increase immediately. Instead, it exhibits a clear plateau (the white arrow after the red dashed line in panel b) during the filament's ascent, followed by a delayed increase in X‑ray flux; post-flare loops appeared about one hour later (see the decaying post-flare loops in panel (b)). The phenomena supports that the torus instability may be the dominant driver of the eruption.}

{We take the electric field $E_y = v_z B_x - v_x B_z$ along the current sheet (CS) in the central x–z cross‑section of the simulation dataset (red line in panels~(c) and (d); its time evolution is available in movie3) as a measure of the reconnection rate~\citep{2000mare.book.....P}. From the red slice that marks the reconnection inflow and outflow regions in panels (c) and (d), we constructed a time–distance diagram (presented in~\nfig{fig6} (e)). The diagram clearly reveals that the magnetic reconnection rate underneath the filament occurs in an intermittent, stepwise manner rather than continuously. In panel (e), we overlay the filament kinematics (green line) together with the timing of the successive jets (red stars), enabling a direct comparison between the filament's ascent and the reconnection rate evolution. One can find that as the filament rises, the reconnection rate of the CS underneath the filament increases in a piecewise manner, mirroring the stepwise increase in the soft X‑ray emission observed in panel~(b). Crucially, the reconnection occurs after jet impact, when the jet impact lifts the filament, the rising filament stretches the background field, subsequently triggering reconnection of the confining magnetic field for the filament. More importantly, this result further indicates that the energy release of a filament eruption can be modulated by the vertical prominence oscillations.}

\section{Conclusions and Discussion}\label{sec:summary}

{Decayless} oscillations are known precursors to filament eruptions \citep{Luna2018,2007SoPh..246...89I,2008A&A...484..487C}. Here, we report a clear event where successive coronal jets impacted a stable filament, excited {transverse and longitudinal} oscillations whose amplitude and period grew as the filament rises, and the filament finally erupted. In order to understand the behavior, we performed MHD numerical simulations to reproduce the oscillatory eruption process driven by successive thermal jets, and analyzed the restoring forces acting on the simulated filament, where the restoring force governing oscillations progressively weakens with the filament rising. Physically, this is analogous to a spring oscillator whose spring constant decreases, naturally leading to the increasing amplitude and period of the filament transverse oscillations. {For the longitudinal oscillations, our model also exhibits some period/amplitude variations; their restoring forces are fundamentally different from those of transverse oscillations, and among the possible controlling factors, the depth of the magnetic dips plays a key role (see the long blue arrow in \nfig{fig6} (c) and (d)).} As the filament rises under the impact of successive jets, it rapidly erupts when the decay index of the background field exceeds the torus instability threshold. Thus, from the observational and simulation evidence, we conclude that the growing oscillations of the filament could be regarded as an obvious precursor for filament eruption.

{In reconnection-favored models (e.g., tether-cutting or shearing-collision models), magnetic reconnection plays a direct driving role: flare loops form early and are concurrent with the initial rise of the filament, as reconnection releases energy and builds the flux rope \citep{2023ApJ...954L..47C,2021NatAs...5.1126J,2025RAA....25e1002L,2001ApJ...552..833M,2013NatPh...9..489S,2003ApJ...585.1073A,2000JGR...105.2375L}.} Our observations and simulations all show the delayed {PFL} during the filament eruption, the filament rises slowly under jet impacts to the torus instability height before any significant {PFL} appear (see the decay PFL in \nfig{fig6} (b)). From our simulation and observational results, the early rise of the filament is facilitated by jet impacts, not by direct reconnection drive. As the filament continued to ascend, the decay index of the background field increases, when it surpassed the torus instability threshold, the torus instability becomes the dominant driver of the eruption, delivering the decisive acceleration that carries the eruption to success and leading to the formation of PFLs. {It should be noted that not all pre‑eruptive filament oscillations are necessarily driven by jets. Other mechanisms, such as mass drainage \citep{Bi2014}, solar surges \citep{2008A&A...484..487C}, nearby flare/jet \citep{2024A&A...691A.354L,2021A&A...646A..12D}, or wave perturbations \citep{2014ApJ...786..151S,2023ApJ...959...71D}, can also induce the growing prominence oscillations and act as facilitators for the filament eruption. }

\section{acknowledgments}
The authors are grateful for the excellent data provided by the SDO, STEREO, NVST, SMART, and Hinode teams. We thank the anonymous reviewer for the constructive comments that greatly improved the manuscript. This work is supported by the Strategic Priority Research Program of the Chinese Academy of Sciences (XDB0560000), the Natural Science Foundation of China (21573005,12303059,12573061), and Yunnan Key Laboratory of the Solar physics and Space Science under No. YNSPCC202212 and No. YNSPCC202506. The numerical computations were conducted on the Yunnan University Astronomy Supercomputer.

\bibliography{ref}
\bibliographystyle{aasjournal}
\clearpage

\end{document}